\documentclass{article}%
\usepackage{amsmath}
\usepackage{amsfonts}
\usepackage{amssymb}
\usepackage{graphicx}%
\providecommand{\U}[1]{\protect\rule{.1in}{.1in}}

\begin{document}
\begin{titlepage}
\vspace{.3cm} \vspace{1cm}
\begin{center}
\baselineskip=16pt \centerline{\Large\bf  Quantum Black Holes} \vspace{2truecm} \centerline{\large\bf Viatcheslav Mukhanov\ \ } \vspace{.5truecm}
\emph{\centerline{Theoretical Physics, Ludwig Maxmillians University,Theresienstr. 37, 80333 Munich, Germany }}
\end{center}
\vspace{2cm}
\begin{center}
{\bf Abstract}
\end{center}
In 1974 Jacob Bekenstein has noticed that the area of the black hole behaves as adiabatic invariant and hence most likely must be quantized
in integers of the Planckian area. In these notes I will present the arguments which led him and myself to such a conclusion and will discuss the consequences of the black hole quantization.
\end{titlepage}

\section{Historical introduction}

In 1985 I was puzzled by the question why the energy (mass) of a black hole in
quantum gravity must be continuous instead of being quantized? In fact, in
many respects a black hole behaves as a bound system, which in quantum theory
has generically a discrete energy spectrum. Assuming that the mass is
quantized I first used an analogy with the hydrogen atom to obtain its
spectrum. The spectrum and the width of the spectral lines for the hydrogen
atom can be derived even without knowing too much about full quantum theory by
just using the Einstein's idea of the quanta of radiation and the
correspondence principle. Treating black hole as \textquotedblleft an
atom\textquotedblright\ which emits typical quanta of the Hawking radiation as
a result of the transitions between quantum levels, I found that the black
hole mass must be proportional to the square root of the integer number and
hence its area is quantized in integers of the Planckian area. Around the same
time Yan Kogan came to a similar conclusion studying \textquotedblleft the
condition for a consistent description of the motion of a test string in the
external field of a black hole\textquotedblright$\cite{Kogan}.$ At the
beginning we were planning to write paper together but finally decided to
split because neither of us had any reasonable explanation for the degeneracy
of quantum levels needed to explain the black hole entropy. First, noting that
the mass spectrum of the excited string behaves similar to the black hole
spectrum we have tried to \textquotedblleft explain\textquotedblright\ the
entropy as originated from the exponential degeneracy of highly excited string
states. Obviously, this did not work straightforwardly because the
corresponding string entropy is proportional to the string mass, while the
black hole entropy must grow as mass squared. An attempt to relate the string
mass to the squared mass of the black hole using gravitational mass defect was
not quite successful. Therefore, I suggested to Yan to publish our findings
separately. In the meantime I decided to try to find more \textquotedblleft
down to earth\textquotedblright\ explanation for the exponential degeneracy of
quantum levels. One of the ideas, widely discussed in the literature, was to
explain the entropy as originated from the different internal geometries of
the black hole, which are indistinguishable from the point of view of an
external observer, for whom this black hole is entirely characterized by its
mass $M,$ angular momentum $J$ and charge $Q.$ To simplify the formulae I will
consider below only the Schwarzschild black hole although the results below
can be easily generalized for the rotating charged black hole. At this time
(not anymore\footnote{In fact, for the given mass $M$ the number of so called
internal geometries must be infinite because, for example, we can first form a
black hole with an arbitrary large mass and then wait until this mass
decreases to $M$ as a result of Hawking evaporation. It is clear that the
internal geometry in this case must be different from those one which we would
have if the black hole was originally formed with mass $M.$ Moreover,
\textquotedblleft inside\textquotedblright\ nonsingular black hole, in
absolute future with respect to the external observer we can have different
kind of universes. The only way to avoid this problem is to relate entropy
with degeneracy due to the unknown way to form the black hole from the matter
of mass $M$.}) I thought that the number of different internal microstates
must be in one to one correspondence with the number of ways to build black
hole of mass $M$. For example, one can produce it out of two photons with
frequencies $\omega_{1}$ and $\omega_{2}$ , satisfying $\omega_{1}+\omega
_{2}=$ $M$ (everywhere I use Planck units, setting $c=\hbar=G=k_{B}=1$).
Another possible way is to form black hole, for example, from three quanta
with frequencies $\omega_{1},\omega_{2}$ and $\omega_{3}$, with $\omega
_{1}+\omega_{2}+\omega_{3}=M$, etc. If no restrictions on the frequencies are
imposed, the number of possible ways to make the black hole is infinite and
hence the corresponding entropy must also be infinite. The only way to rescue
the entropy definition above is to assume that the black holes are quantized.
This imposes constrains on the frequencies making the number of ways to form
black hole to be finite. These arguments have led me to the same quantization
law I obtained before. These two completely different ways of reasonings for
the area quantizations I published first in Lebedev Institute preprint
$\cite{M1}$. However, in the published version of the paper $\cite{M2}$ I
skipped the derivation by analogy with the hydrogen because of rather severe
restrictions on the number of pages in JEPT letters. The\ preprint I have sent
to the only person outside of the Soviet Union, to Jacob Bekenstein. Leaving
inside of, as I thought, \textquotedblleft eternal black
hole\textquotedblright\ called Soviet Union\ I did not care too much about
making my results known to the \textquotedblleft external
observers\textquotedblright. However, I could not resist sending paper to
someone who originated the whole concept. Moreover, Ya. Zeldovich was strongly
opposing this idea claiming that the levels will be spitted and degeneracy
completely removed as it happens for usual thermodynamical systems in
equilibrium with the thermal bath. Although I tried to argue that the black
hole cannot be treated as a normal thermodynamical system because of its
negative heat capacity, Zeldovich was insisting that my arguments are silly.
To my great surprise I got an answer from Jacob, where he wrote that he
completely agrees with me on all points including the degeneracy issue. He
also pointed out that he came to the conclusion that the area must be
quantized long ago basing on observation that it behaves like adiabatic
invariant $\cite{B1}.$ I was extremely happy. This meant that I am not alone
and there is another person in the world who also likes this \textquotedblleft
silly\textquotedblright\ idea.

After collapse of Soviet Union I went to ETH (Zurich) and in 1993 wrote to
Jacob \textquotedblleft inviting myself\textquotedblright\ to Jerusalem. His
reaction was positive and for the first time in my life I went to the homeland
of all religions where I finally met Jacob personally. Someone who has already
overlapped with him before told me that Jacob is extremely religious and not
very \textquotedblleft communicative\textquotedblright. This impression
happened to be completely wrong. First of all Jacob was just moderately
religious, perhaps a bit selective in a choice of discussion partners, but I
did not find on his side even a slightest trace of self-conceit or, so wide
spread in our community, \textquotedblleft inferiority or superiority
complexes\textquotedblright\ (often hardly distinguishable from outside). One
could feel immediately his strong character. Jacob never cared too much about
\textquotedblleft main stream\textquotedblright\ research or sociology and was
entirely occupied just with those ideas which he was considering to be
interesting irrespective of the \textquotedblleft community
fashion\textquotedblright. His judgement of people was rather sharp, but he
was keeping it mainly for himself and only in rare cases was sharing with the
close friends. Jacob's opinions on questions of science and politics might be
sounded not always \textquotedblleft politically correct\textquotedblright\ in
a modern useless sense of this notion, but were exactly to the point and very
deep. He was a person of value and cared more about the essence than about
making impression on outside. In spite of the fact that sometimes Jacob looked
as a very \textquotedblleft soft person\textquotedblright\ he could fight to
the bitter end if he thought that something is wrong and worth fighting. This
does not mean at all that he was stubborn and I witnessed how in couple of
cases Jacob changed his mind without any problem just under the pressure of
facts. Many of these observations I made much later but from the very
beginning we found common language without any difficulties (might be because
on many relevant issues we were sharing the same opinion). Therefore my
interactions with Jacob were always extremely friendly and fruitful and in a
short time we became real friends.

I was invited to Jacob's place and introduced to his wife Bilha and kids,
Yehonadav, Uriya and Rivka, who by now have grown to the worthy people and I
recall that Jacob was always proud of them. Although during my first visit to
Jerusalem my touristic preferences prevailed over scientific interests we,
nevertheless, found time for discussions and as a result in a year have
published the paper with detailed derivation of the quantum black hole
spectrum $\cite{BM}.$ Working on this paper and facing problems I was ready
(many times) to give up. However, it was not so easy to do so working with
Jacob. Finally, I realized what Einstein had really meant when he wrote
\textquotedblleft It is not that I am so smart, it is just that I stay with
problems longer\textquotedblright. Having this experience with Jacob and after
I head his story how he got an idea about black hole entropy I realized that
this Einstein's remark is entirely applicable to Jacob. Being on the track he
was infinitely persistent, might be not always extremely quick, but was never
giving up if he had a feeling that there remain questions to be understood and
clarified. Jacob intuition was impressive and his understanding of statistical
physics was the deepest one I ever met.

After our paper was finished we started to look for an operator algebra which
would allow us to formalize the description of quantum black holes. With this
purpose Jacob visited me in Zurich. I was impressed when after discussing the
properties of the algebra, Jacob mentioned that Bryce DeWitt told him that
everything about algebra is in Jacobi identities, went to his office and soon
was back with Virasoro algebra about which he has completely forgotten (if
knew) and rediscovered by himself with a half an hour. For obvious reasons
this algebra did not work because degeneracies were not large enough. Another
interesting memory to some extent characterizes the Jacob's attitude about
life preferences. I recall that when someone complained that during two weeks
he could not find enough time to talk to Jacob because was too busy, Jacob
replied with unforgettable ironic smile under his mustache \textquotedblleft
If somebody is too busy this just means that he is doing too many irrelevant
things\textquotedblright. Sometimes he could be really sharp.

Since I moved to Germany in 1998, our worldlines were often crossing either in
Jerusalem or on some \textquotedblleft neutral territory\textquotedblright%
\ as, for instance, in IHES in Bures-sur-Yvette or in Princeton. In
particular, in 1998 we have continued the search for the right algebra during
our visit to IHES. Unfortunately, without final success. These attempts Jacob
has summarized in $\cite{B3}$ and $\cite{B4}.$ Every time when I was coming to
Jerusalem we were discussing all possible questions from science to politics
and life. Jacob was an exceptional human. Several times when I had really
serious problems in my life only Jacob, as nobody else, could find absolutely
right words to support me. Last time we met in Jerusalem Winter School in 2015
and arranged our next meeting to IHES in October. Just a couple of months
before Jacob passed away...

Below I describe simple reasonings in favor of the black hole quantization.
First, I will present never before published derivation of the black hole
spectrum basing on analogy with the hydrogen atom. Forgetting about quantum
mechanics I will consider the simplest model for the classical atom and with
the minimal assumption about quanta of radiation will show how the Bohr
spectrum can be derived. Moreover just using correspondence principle I will
estimate the width of the levels and show that the result obtained is in
agreement with full quantum theory. We will see that even for the highly
excited states the classical physics is not reproduced precisely in apparent
conflict with naive interpretation of the correspondence principle. I hope
that this consideration will convince reader that the derived by analogy black
hole quantization with its unexpected consequences for the Hawking radiation
might make sense in spite of the fact that we do not have yet a consistent
theory of non-perturbative quantum gravity. In the remaining part of the paper
I will recall the justification for the entropy of quantum black hole and show
that the modified thermal spectrum is just a consequence of the exponential
degeneracy of quantum levels.

\section{Quantization}

\textbf{Hydrogen.} Let us consider the hydrogen atom and verify whether we can
derive its energy levels and their width only by using a) the Einstein quanta
of radiation and b) the correspondence principle. The quantized energy $E$
depends on the integer number $n$ and our task is to determine $E_{n}=E(n)$
neglecting all other quantum numbers, such as angular momentum and
spins$.$Moreover we simplify the classical picture assuming that the electron
is moving on circle orbit of radius $r$ around the proton at rest$.$ According
to Einstein the quantum transition from $n$-th to $n-1$ level must be
accompanied by the emission of photon with frequency
\begin{equation}
\omega_{n,n-1}=E_{n}-E_{n-1}. \label{1}%
\end{equation}
In its minimal interpretation the correspondence principle would mean that for
large $n$ this frequency must coincide with the frequency of classical
radiation which is about the frequency of rotation of the electron around
center. On a bound orbit of radius $r$ the kinetic energy of the electron,
$T=mv^{2}/2,$ is twice less than the magnitude of the negative potential
energy $U=-\alpha/r,$ where $\alpha=e^{2}$ is the fine structure constant.
Hence, the total energy is%
\begin{equation}
E=T+U=-\frac{\alpha}{2r}=-\frac{m\omega^{2}r^{2}}{2}, \label{2}%
\end{equation}
from where it follows that%
\begin{equation}
\omega=\left(  \frac{\alpha}{mr^{3}}\right)  ^{1/2}=\left(  -\frac{8}%
{m\alpha^{2}}E^{3}\right)  ^{1/2}. \label{3}%
\end{equation}
Taking into account that $E_{n}-E_{n-1}\simeq dE_{n}/dn$ for $n\gg1$ and
equating $\omega_{n,n-1}$ from $\left(  \ref{1}\right)  $ to $\omega$ we
obtain the following equation for $E_{n},$
\begin{equation}
\frac{dE_{n}}{dn}\simeq\left(  -\frac{8}{m\alpha^{2}}E_{n}^{3}\right)  ^{1/2},
\label{4}%
\end{equation}
solving which gives us the spectrum%
\begin{equation}
E_{n}=-\frac{m\alpha^{2}}{2n^{2}}, \label{5}%
\end{equation}
which is valid even for small $n$ although the derivation above is justified
only for $n\gg1$. For large $n$ the distance between the nearby levels is
about%
\begin{equation}
\Delta E_{n}=E_{n}-E_{n-1}\simeq\frac{m\alpha^{2}}{n^{3}}\simeq\omega_{n}.
\label{6}%
\end{equation}
One could wonder whether for $n\gg1$ we must obtain exactly classical behavior
or there will be still essential differences between the classical and quantum
hydrogen even in this case? To answer this question we must determine the
width of the highly excited levels and compare it with the distance between
the nearby levels. With this purpose let us use the correspondence principle
in its minimal formulation by assuming that the overall intensity of the
emitted radiation by quantum atom must be in agreement with the result
obtained in the classical theory. The classical radius of the electron orbit
corresponding to energy $E_{n}=-\alpha/2r_{n},$ given in (\ref{5}), is%
\begin{equation}
r_{n}=\frac{n^{2}}{m\alpha}. \label{7}%
\end{equation}
According to the Maxwell theory the intensity of the emitted radiation is
given by the second time derivative of the dipole momentum, that is,%
\begin{equation}
I\simeq\left(  e\ddot{r}\right)  ^{2}\simeq\alpha\omega^{4}r^{2}, \label{8}%
\end{equation}
or substituting here $\omega$ and $r$ from $\left(  \ref{6}\right)  $ and
$\left(  \ref{7}\right)  \ $we obtain%
\begin{equation}
I_{n}\simeq\frac{m^{2}\alpha^{7}}{n^{8}}. \label{9}%
\end{equation}
Thus, the time interval $\tau_{n}$ during which atom emits the energy needed
for transition from $n$ to $n-1$ levels, given in $\left(  \ref{6}\right)  ,$
is equal to%
\begin{equation}
\tau_{n}\simeq\frac{\Delta E_{n}}{I_{n}}\simeq\frac{n^{5}}{m\alpha^{5}}%
\simeq\frac{n^{2}}{\alpha^{3}}\omega_{n}^{-1}. \label{10}%
\end{equation}
Correspondingly the width of the level $n$ is%
\begin{equation}
W_{n}\simeq\frac{1}{\tau_{n}}\simeq\frac{m\alpha^{5}}{n^{5}}\simeq\frac
{\alpha^{3}}{n^{2}}\Delta E_{n}, \label{11}%
\end{equation}
and it gets smaller compared to the distance between nearby levels as $n$
grows. Hence the emitted lines of the isolated hydrogen atom become sharper
and the quantum nature of the hydrogen become even more distinctive as $n$
grows. The detailed picture of emission by the quantized atom is very
different from the classical picture. According to classical theory the
transverse part of the vector potential $A$ at distance $R\simeq\omega
_{n}^{-1}$ is%
\begin{equation}
A=\frac{\dot{d}}{R}\simeq\frac{e\omega_{n}r_{n}}{R}\simeq\frac{e^{3}}{n}%
\frac{1}{R}, \label{12}%
\end{equation}
and by factor $e^{3}/n$ smaller than the level of inevitable quantum
fluctuations in scales $R$ in conflict with uncertainty relation for
radiation. According to classical theory the electron continuously emits
radiation during the time interval $\tau_{n}$ changing energy from $E_{n}$ to
$E_{n-1}.$ In quantum theory a quantum of radiation of frequency $\omega_{n}$
is emitted randomly during the time interval $\omega_{n}^{-1}$ per
$n^{2}/\alpha^{3}$ rotations. From the formulae above we see that the quantum
features of the isolated hydrogen atom become more distinctive for large $n.$
For instance, in a hypothetical case $e\gg1$ the discrete energy levels make
sense only for $n>e^{3}$ because otherwise their width is larger than the
distance between the nearby levels.

\textbf{Black Hole.} Now I will apply a similar line of reasoning for a black
hole of mass $M$ assuming that the mass is a function of the integer number
$n$. It is not excluded that this assumption might find a justification in
some unknown yet non-perturbative quantum gravity. However, in the absence of
such a theory we can only try to guess what could be the spectrum of the black
hole in a way similar to how we did it above for the hydrogen atom. Although
the consideration below can be easily generalized for a charged, rotating
black hole \cite{M2} to simplify the formulae I will consider here only the
Schwarzschild black hole, which from the point of view of external observer is
even more simple than hydrogen because it is entirely characterized only by
its mass. Once again I will use the correspondence principle and the quanta of
Hawking radiation (instead of Einstein's photons). For quantum black hole this
radiation can be treated as spontaneous radiation due to the transition from
$n$ to one of the nearby levels, for instance, to $n-1$ level$.$ The typical
frequency of the Hawking quanta is of order of black hole temperature,
\begin{equation}
T_{H}=\frac{1}{8\pi M}.\label{12a}%
\end{equation}
According to the correspondence principle, for large $n$%
\begin{equation}
\frac{dM_{n}}{dn}\simeq M_{n}-M_{n-1}\simeq\gamma T_{H},\label{13}%
\end{equation}
where $\gamma$ is a numerical coefficient of order unity, and hence
\begin{equation}
\frac{dM_{n}}{dn}\simeq\gamma\frac{1}{8\pi M_{n}}.\label{14}%
\end{equation}
Integrating this equation we obtain the mass spectrum
\begin{equation}
M_{n}\simeq\sqrt{\frac{\gamma}{4\pi}n},\label{15}%
\end{equation}
and therefore the area of black hole $A_{n}$ must be quantized in terms of
integers of the Planck area,%
\begin{equation}
A_{n}=16\pi M_{n}^{2}=4\gamma n.\label{16}%
\end{equation}
We will show below that the numerical coefficient $\gamma$ in $\left(
\ref{13}\right)  $ is most likely equal to $\ln2$ and hence the distance
between the nearby levels is
\begin{equation}
\Delta M_{n}=M_{n}-M_{n-1}\simeq\frac{\ln2}{8\pi M_{n}}.\label{17}%
\end{equation}
To estimate the width of the levels we use the correspondence principle
assuming that the total flux of Hawking radiation from the classical black
hole must be about the same as the flux from quantum black hole (similar how
we did it for the hydrogen). According to \cite{DEWITT}%
\begin{equation}
\frac{dM}{dt}\simeq-\frac{9N}{10\pi\cdot8^{4}\cdot M^{2}},\label{18}%
\end{equation}
where $N$ is the number of distinct massless quanta in Nature. On the other
hand the rate of emission for the highly excited quantum black hole is%
\begin{equation}
\frac{dM_{n}}{dt}\simeq-\frac{2\Delta M_{n}}{\tau_{n}}\simeq-\frac{2\ln2}{8\pi
M_{n}\tau_{n}},\label{19}%
\end{equation}
where $\tau_{n}$ is the lifetime on level $n.$ The factor $2$ here accounts
for the possibility of direct transitions to $n-2,$ $n-3$ et.cet. levels
\cite{BM}. Comparing (\ref{18}) and (\ref{19}) we find
\begin{equation}
\tau_{n}\simeq\frac{5\cdot8^{4}\cdot\ln2\cdot M_{n}}{18N},\label{20}%
\end{equation}
and correspondingly the width of $n$-th level is%
\begin{equation}
W_{n}\simeq\frac{1}{\tau_{n}}\simeq\frac{18N}{5\cdot8^{4}\cdot\ln2\cdot M_{n}%
}\simeq4.5\cdot10^{-2}N\text{ }\Delta M_{n}\label{21}%
\end{equation}
For $N<20$ the level width is less than the distance to nearby level.
Therefore, the spectrum of the quantum black hole can be very distinctive from
the usual Planckian spectrum even for the large black holes. As it was shown
in \cite{BM}, the resulting spectrum is perfectly consistent with the thermal
spectrum of photons in a finite box with reflecting boundaries. In this latter
case the frequencies are proportional to the integers of the inverse size of
the box and the spectrum is not continuous but consists of lines with
intensities obeying the Planckian distribution. Similar for the black holes
instead of continuous spectrum we have the thermal spectrum consisting of
rather sharp lines.

Thus, we have found that the spectrum of radiation from the large quantum
black holes can be very different from the continuous Hawking spectrum which
he found considering quantum fields in a given classical background. It is not
as surprising as it looks at the first glance if we recall the situation with
the highly excited hydrogen states where the line structure of emitted
radiation becomes even more profound for large $n.$ Therefore, the
correspondence principle in both cases above should be understood in a rather
restrictive way. Namely, at very large $n$ only the averaged characteristic of
the emitted radiation, but not its detailed features, must be in agreement
with the classical theory. If the consideration above is correct, this opens
experimental possibility to check the effects of unknown yet nonperturbative
quantum gravity observing radiation from the exploding primordial black holes.

\subsection{Entropy}

To define the statistical entropy of quantized black hole we assume that it is
due to the lost information about the way how the black hole at level $n$ was
formed. Because of the event horizon this information can never be recovered
by the external observer. The number of different ways to form the quantum
black hole from a given matter is finite. For example, one possible way is
first to form a smallest black hole at the first level and after that go step
by step to the level $n$. The other possibility is to jump directly to level
$n$ without intermediate steps et. cet. The total number of ways to get to the
level $n$ is obviously equal to the number of subdivisions of the integer
number $n$ into ordered sums (for instance, for $n=3$ we have
$3=1+1+1,3=1+2,3=2+1,3=3$ where one has to distinguish $1+2$ from $2+1$). The
result for the level $n$ is
\begin{equation}
\Gamma\left(  n\right)  =2^{n-1}. \label{22}%
\end{equation}
and the black hole entropy is equal
\begin{equation}
S_{n}=\ln\Gamma\left(  n\right)  =\left(  n-1\right)  \ln2. \label{23}%
\end{equation}
It is proportional to the \textquotedblleft elementary\textquotedblright%
\ entropy $\ln2$, corresponding to one bit of information. Assuming that the
relation between entropy and area
\begin{equation}
S=\frac{1}{4}A+const, \label{24}%
\end{equation}
still holds we obtain the following quantization rule
\begin{equation}
A_{n}=4\ln2\cdot n, \label{25}%
\end{equation}
in full agreement with $\left(  \ref{16}\right)  $ if we set $\gamma=\ln2$ and
take the constant of integration in $\left(  \ref{24}\right)  $ to be equal
$-\ln2,$ so that $n=0$ describes the state without black hole. The minimal
increase of the area, $\Delta A=4\ln2,$ is in agreement with the earlier
result by Bekenstein \cite{B1} (see also \cite{DEWITT}) and it was obtained
before he suggested the area quantization in \cite{B2}.

As it follows from $\left(  \ref{22}\right)  $ each level of the black hole is
exponentially degenerate. At the first glance one could expect the splitting
of these highly degenerate levels after black hole gets in equilibrium with an
infinite thermal bath which we need to define its macrocanonical entropy.
However, the black hole has a negative heat capacity and therefore
macrocanonical definition is not applicable here. One has to use the
microcanonical ensemble instead. In this case there is no any reason to expect
the splitting of the levels. Moreover, as it was shown in \cite{BM} the
perfectly thermal properties of the radiation are entirely due to the
exponential degeneracy of the levels. To give an idea why this happens let us
consider the large black hole at level $n\gg1.$ Then all energy levels between
$n$ and $n-m$ for $m\ll n$ are nearly equidistant with the energy difference
between nearby levels corresponding to the fundamental frequency%
\begin{equation}
\bar{\omega}=\frac{\ln2}{8\pi M_{n}}=\ln2\cdot T_{H}. \label{26}%
\end{equation}
It is clear that the black hole can directly go from the state $n$ to any
nearby lower states and, for instance, going to the $n-2$ level emits quantum
of radiation with the frequency $2\bar{\omega}$ or with the frequency
$\omega=m\bar{\omega}$ for the direct transition from $n$-th to $n-m$ level.
For the external observer the black hole is even more simple than hydrogen
atom because it has no hairs and entirely characterized by its mass and
entropy. Therefore one can expect that the matrix elements for all transitions
to the nearby levels are the same and the probability for direct transition
from $n$ to $n-m$ level with emission of quantum of frequency $\omega
=m\bar{\omega}$ is simply proportional to the degeneracy of $n-m$ level, that
is,%
\begin{equation}
P(\omega)\propto2^{n-m}\propto\exp\left(  -\ln2\cdot\frac{\omega}{\bar{\omega
}}\right)  =\exp\left(  -\frac{\omega}{T_{H}}\right)  . \label{27}%
\end{equation}
It was shown in $\left(  \cite{BM}\right)  $ that probabilities for all
possible sequences of quanta are precisely as for the thermal radiation with
temperature $T_{H}$ in a finite box of size $\bar{\omega}^{-1}.$ It is rather
remarkable that all thermal properties of the Hawking radiation follow from
the degeneracy (entropy) of quantized black hole under assumption of equal
matrix element for the transitions to close levels. Although the spectrum of
the quantum black holes is thermal it nevertheless different from the
continuous spectrum derived by Hawking. It must consist of lines with the
width of few percent of the distance between nearby lines. Moreover, the
minimal frequency which black hole can emit is $\bar{\omega}=\ln2\cdot T_{H}.$
The probability of the emission of quanta with the wavelength much larger than
the size of the black hole is strongly suppressed.


\bigskip

\begin{thebibliography}{9}                                                                                                %


\bibitem {B1}J. Bekenstein, \textit{Generalized second law of thermodynamics
in black-hole physics, }Phys.Rev. D, v.9, No. 12, p. 3292, 1974.

\bibitem {B2}J. Bekenstein, \textit{The Quantum Mass Spectrum of the Kerr
Black Hole, }Lett. al Nuovo Cimento. v.11, No. 9, p. 467, 1974.

\bibitem {BM}J. Bekenstein, V. Mukhanov, \textit{Spectroscopy of the quantum
black hole}, Phys.Lett. B360, p. 7, 1995.

\bibitem {B3}J. Bekenstein, \textit{Quantum Black Holes as Atoms, }in
Proceeding of 8th Marcel Grossman Meeting, Jerusalem, Israel, 1997, Editors T.
Piran and R. Ruffini, World Scientific,1999.

\bibitem {B4}J. Bekenstein, \textit{The Case for Discrete Energy Levels of a
Black Hole, }Int.J.Mod.Phys., A17S1, p. 21, 2002.

\bibitem {DEWITT}B. DeWitt, \textit{Quantum Field Theory in Curved Space,
}Phys. Rep., v.19, No. 6, p. 296, 1975.

\bibitem {Kogan}Ya. Kogan, \textit{Qantization of the mass of a black hole in
string theory, }JETP. Lett. v.44, No. 4, p. 267, 1986.

\bibitem {M1}V. Mukhanov, \textit{Evaporation and Entropy of Quantized Black
Hole, }Lebedev Institute preprint N 163, 1986

\bibitem {M2}V. Mukhanov, \textit{Are Black Holes Quantized?, }JETP. Lett.
v.44, No. 1, p. 63, 1986.
\end{thebibliography}
\end{document}